# Software Implementation And Evaluation Of Lightweight Symmetric Block Ciphers Of The Energy Perspectives And Memory

[1]Jaber Hosseinzadeh, [2]Abbas Ghaemi Bafghi

[1]Data and Communication Security Laboratory (DCSL), Faculty of Engineering, Ferdowsi University of Mashhad

[1]Jaber_hosseinzadeh@mail.um.ac.ir

[2]Data and Communication Security Laboratory (DCSL), Department of Computer Engineering, Faculty of Engineering, Ferdowsi University of Mashhad

[2]Ghaemib@um.ac.ir

**Abstract**

Lightweight ciphers are the form of encryption that strictly limited to devices such as tags, RFID, wireless sensor networks applications. Low-resource devices has many limitations in power, energy and memory. In this work, the lightweight block ciphers is implemented on the Atmega128 microprocessor and the results of the energy perspectives and memory were assessed. The results of the evaluation show that the SPECK(64,96) cipher has been the best value of the perspective of energy and is appropriate for wireless sensor networks with the main requirement of energy. If the measure is memory usage, the TWINE-80 is the most appropriate implemented lightweight ciphers on the Atmega128 microprocessor. The SPECK(64,96) cipher is in third place of the perspective of consumer memory. In other side, TWINE-80 cipher is In the second place of the perspective of energy. SPECK(64,96) and TWINE-80 ciphers are the most appropriate for wireless sensor networks.

1- Introduction

Lightweight cryptography is considered as a type of cryptography that is used for strictly constrained devices such as RFID tags and wireless sensor networks. Resource constrained devices have many limitations in the power, energy and memory. These devices are often used in critical and difficult access (such as military) environments and deal with limited battery on insecure communication channels. These factors suggest strong cryptography solutions for these devices. Conventional cryptography algorithms such as RSA and AES consume lots of power and energy and provide more safety in these systems. Hence, we need new ways of cryptography which provide the restrictions of the resource constrained devices. This method is called lightweight cryptography [1, 6, 9]. In general, cryptography is divided into the two categories: symmetric and asymmetric. In symmetric cryptography, the same key number is used for both encryption and decryption. But in asymmetric cryptography, two different but related keys are used to encrypt and decrypt the codes [7,10]. In cryptography operation, single key is used as public key which is available for everyone. The private key is used for decryption. Lightweight symmetric ciphers are divided into two categories: block and stream [1]. In this paper, some symmetric lightweight block ciphers are implemented and also evaluated using energy criteria. A brief description of symmetric lightweight block ciphers is provided in the following.

2- Implementation of lightweight block ciphers

Lightweight block ciphers including KLEIN-80, TWINE-80, Piccolo-80, SPECK (64,96) and SIMON(64,96) are implemented on the Atmega128 processor in simulation environment of AVR studio 5.1 which is similar to visual studio 2008 software. The simulation codes are based on C and the compiler translates the codes and simulate on Atmega128 processor. There are two basic steps including target device selection and simulation tools to work with the software. Processors target devices are Atmega128 and simulation tool is AVR simulator. After writing codes and pressing "start debugging and break" button, the codes are run and break wherever there is break point so that the intermediate results can be considered. To pass the breakpoint, user must push manually the green triangle-shaped button named "continue".

3- Evaluating lightweight ciphers based on energy criteria

Energy is one of the main constraints in resource-constrained devices. This criterion is major and important for wireless sensor networks. The energy can be modeled for wireless sensor networks [8], [11]. The energy can be calculated by:

$$Energy = \frac{P(W) * Clock\ cycles}{Frequency(hz)} \quad (1)$$

In fact, the power can be obtained using the production of voltage and current. The Joule is the energy unit. In the above equation, frequency indicates CPU operating frequency in hertz and P is the consumed power. Currents and voltages are considered in Ampere and Volt, respectively. Clock cycles show the number of cycles that the processor needs to execute the codes. In energy, the numbers of cycles are achieved by simulator environment. Current and voltage are obtained based on active mode values.

Table1- the energy of the implemented lightweight ciphers on atmega128 processors in Micro-Jules

| Cipher | Clock cycles | Energy(micro joul) |
|---|---|---|
| KLEIN-80 | 17244 | 86.22 |
| TWINE-80 | 14242 | 71.21 |
| Piccolo-80 | 16748 | 83.74 |
| SIMON(64,96) | 19130 | 95.65 |
| SPECK(64,96) | 13868 | 69.34 |

As we can see in table 1, in implemented lightweight ciphers, SPECK (64,96) has the best value in terms of energy. TWINE-80 and Piccolo-80 are next best ranks of lightweight ciphers in terms of energy. Figure 1 shows the energy of the implemented lightweight ciphers.

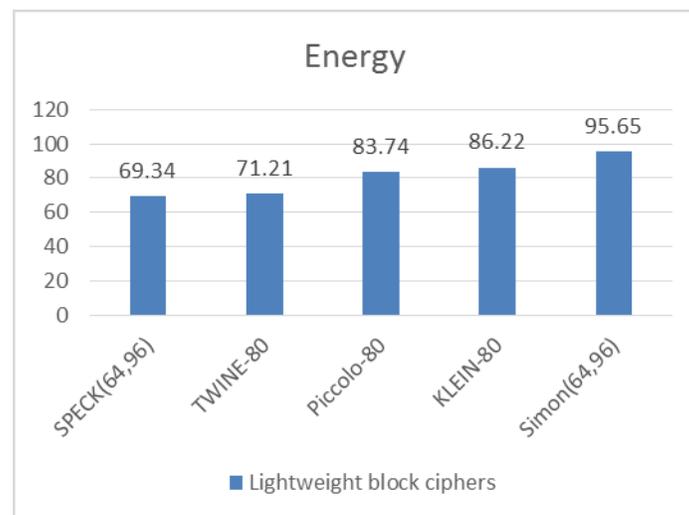

Figure1- the energy of the implemented lightweight ciphers on atmega128 processors in Micro-Jules

## 4- Evaluating lightweight ciphers based on memory criteria

A memory criterion represents the device memory usage when we use cipher in our algorithms. Memory requirements can be divided into the two sections: FLASH and RAM. Implemented lightweight ciphers are evaluated In terms of memory usage.

Table 2- FLASH and RAM memory required for ciphers implementation on Atmega128

| Cipher | RAM(bytes) | FLASH(bytes) |
|---|---|---|
| KLEIN-80 | 76 | 1178 |
| TWINE-80 | 95 | 1140 |

| Piccolo-80 | 76 | 1691 |
| --- | --- | --- |
| SIMON(64,96) | 168 | 1406 |
| SPECK(64,96) | 126 | 1391 |

RAM memory and FLASH memory usage are presented in figure 2 and 3, respectively.

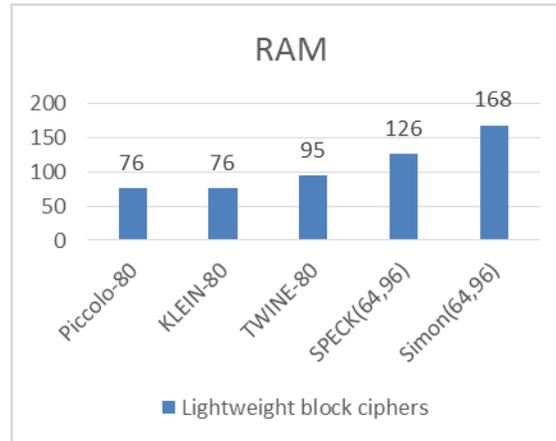

**Figure 2- The amount of RAM usage for implemented lightweight ciphers on the Atmega128 CPU.**

As we can see by figure 2, Piccolo-80 and KLEIN-80 has the least amount of RAM usage, and TWINE-80 and SPECK (64,96) ciphers are next in rank.

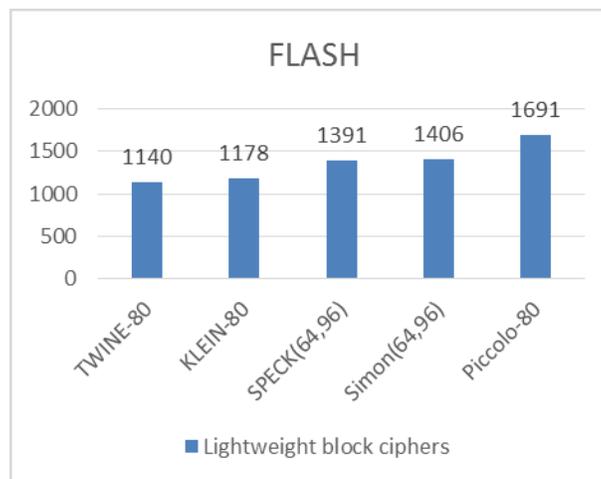

**Figure 3- The amount of FLASH usage for implemented lightweight ciphers on the Atmega128 CPU.**

As we can see by figure 3, TWINE-80 cipher has the least amounts of FLASH usage. KLEIN-80 and SPECK (64,96) ciphers are next in rank.

For better comparison of implemented lightweight ciphers in terms of memory usage, we use both RAM and FLASH simultaneously as the public memory. Figure 4 are derived from a combination of the figure 2 and 3. Figure 4 evaluates the implemented lightweight ciphers in terms of memory usage. The values of these two tables are obtained from the combination of figures 2 and 3 values.

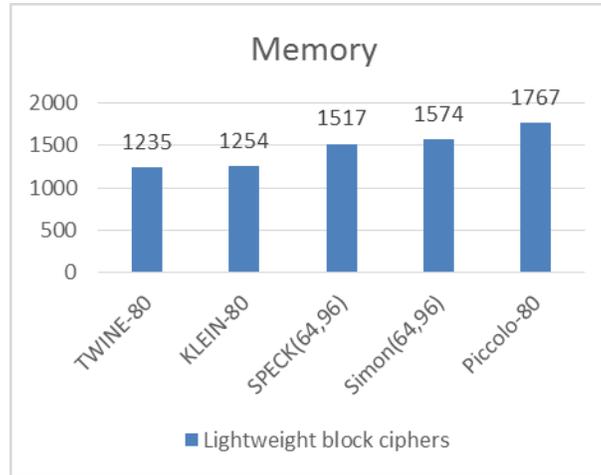

**Figure 4- The amount of memory usage for implemented lightweight ciphers on Atmega128 processor.**

As we can see by figure 4, TWINE-80 cipher has the least amounts of FLASH usage. KLEIN-80 and SPECK (64,96) ciphers are next in rank. Therefore, TWINE-80 cipher is the best lightweight cipher for Atmega128 processor in terms of memory usage.

**5- Conclusion**

In this paper, lightweight ciphers are implemented on Atmega128 processor and evaluated based on the energy and memory criteria. Wireless sensor networks are considered as the main benchmark. SPECK (64,96) has the best value in terms of energy and the best lightweight block ciphers for WSNs. The best lightweight block ciphers in terms of memory usage, is TWINE-80. SPECK (64,96) cipher is in the third rank of the best lightweight ciphers. The results also states that the TWINE-80 cipher is in the second rank in terms of energy criteria. So we can consider both SPECK (64,96) and TWINE-80 as the best lightweight cipher to use in WSN.